\documentstyle[12pt,epsf]{article}

\begin{document}
\newcommand{\preprint}[1]{\begin{table}[t]  
           \begin{flushright}               
           \begin{large}{#1}\end{large}     
           \end{flushright}                 
           \end{table}}                     
\baselineskip 22pt
\preprint{TAUP-2455-97}

\title{\bf Is the Planck Momentum Attainable?}
\author{\bf Aharon Casher\thanks{Email:ronyc@post.tau.ac.il} and Shmuel
  Nussinov \thanks{Email:nussinov@post.tau.ac.il}\\ 
\em Raymond and Beverly Sackler Faculty of Exact
 Sciences\\ \em School of Physics and Astronomy\\ 
\em Tel Aviv University,Ramat Aviv, 69978, Israel
\\
 \em and 
\\
\em Department of Physics and Astronomy \\\em University of South Carolina\\
\em Columbia , SC , USA}

\maketitle

\maketitle
\begin{abstract}
  
 We present evidence that an interplay of the laws of microphysics and
cosmology renders the Planck momentum unattainable by an elementary
particle. Several categories of accelerators are analyzed and shown to fail.

\end{abstract}

\newpage
\section{Introduction}
The Planck mass, $m_{Pl} = (\hbar c/G_{Newton})^{1/2} \approx 10^{19}\; GeV$ and
corresponding length $l_{Pl} = \hbar/m_{Pl}c \approx 10^{-33} cm$  or time $t_{Pl} =
l_{Pl}/c$, are of fundamental importance, marking the onset of strong
non-renormalizable quantum gravity effects.
In this ``superplanckian'' regime the theoretical framework of local field theory 
and indeed the very concept of space time may break down. Many different lines 
of reasoning suggest that $l_{Pl}$ and $m_{Pl}$ are the minimal distance to which the 
location of an elementary particle can be defined and the maximal energy to which 
an elementary localized degree of freedom can be excited. 

In the following we 
address the question of the highest energy that can be given to a single 
elementary particle in our universe subject to the known laws of microphysics 
and cosmology. As one approaches the Planck regime novel physics effects may come 
into play, modifying the very concept of an elementary degree of freedom. This 
new, fundamental, Planck scale physics may, in turn, directly prevent crossing the 
``Planck barrier'', i.e. prevents achieving $W >> m_{Pl}$ for one elementary degree of freedom. 
However we are starting with the familiar low energy regime where weakly coupled 
theories, with degrees of freedom corresponding to local fields or pointlike quarks, 
leptons, photons, and other gauge bosons apply. Our question whether any device 
utilizing electromagnetism, gravity, or strong interaction (QCD) can accelerate 
such an elementary particle to an energy $W \geq m_{Pl} \approx 10^{19} GeV$ is therefore well 
posed. Interestingly we find that an interplay of microphysics and cosmological 
parameters may prevent acceleration to $W_{Pl} \approx 10^{19} GeV$ \underline{already 
at this level}. 

Strictly speaking accelerating one particle to $W=m_{Pl}$ causes 
only one, longitudinal, ($z$) dimension to shrink up to $\Delta z \leq l_{Pl}$. It 
does not, by itself, generate say energy densities in the superplanckian regime: 
$U\equiv \frac{dW}{dV}\geq {m_{Pl}}^4$. To achieve 
the latter in say a $qq$ collision\footnote{In reality $qq$ scattering is generated 
via $pp$ collisions. The momentum fraction carried by the valence quarks 
evolves to zero as $Q^2 \rightarrow \infty$ but only logarithmically. Thus we can still 
achieve in principle the superplanckian $qq$ collisions if the protons are 
accelerated to somewhat higher energies $m_{Pl}* ln(m_{Pl}/\Lambda_{QCD})$.} we need to accelerate 
the two quarks in opposite directions so as to have an invariant center of mass 
energy $\sqrt{(P_1 + P_2)^2} \approx m_{Pl}$ and then both have to mutually scatter 
with a transverse momentum transfer $\left | \vec q \right | \approx m_{Pl}$ 
so as to achieve localization also in the transverse (x,y) directions down to 
$\Delta x, \Delta y \leq l_{Pl}$. The cross section for such a collision, 
$\sigma \approx {l_{Pl}}^2$, is extremely small $(\approx 10^{-66} cm^2)$ 
making the goal of achieving superplanckian energy densities far more difficult 
than the mere acceleration of one particle to $W \geq m_{Pl}$.\footnote{Naively one 
would think that if an accelerator capable of accelerating one proton to 
$m_{Planck}$ can be built, there is no intrinsic difficulty in joining two such 
accelerators back to back to achieve CMS invariant Planck energy.} We find 
however that even the latter goal seems 
to be unattainable. It should be emphasized that this does not stem from any 
\underline{kinematic} limitation of say a maximal Lorentz boost. Thus boosts 
vastly exceeding $\gamma = 10^{19}$ (or $2.10^{22}$) required to achieve 
Planck energy proton (or electron) are implicit in having very energetic 
photons. The latter can be viewed as very soft photons boosted by 
$\gamma >> 10^{23}$.

The group property of Lorentz boosts implies that one very large boost can be achieved 
in many successive small steps. Thus consider the set of ``Gedanken'' nested accelerators 
illustrated schematically in Fig. (1). Suppose that a proton is accelerated inside some 
microscopic 
device $D_1$(see
Fig. (1)) to $\gamma = 11$.  The whole $D_1$ device in turn sits within a
larger setup $D_2$, which boosts $D_1$ to $\gamma = 11$.  $D_2$ in turn sits
inside $D_3$... etc., etc.  The device $D_{18}$ is then also boosted by
$D_{19}$ to $\gamma = 11$, thus finally achieving transplanckian energies $E
= \gamma^{19}\, GeV \approx\, 10^{20}\, GeV$ for our proton. 
It is hard to imagine a non-local mechanism impeding the operation of the huge accelerator 
$D_{19}$ just because nineteen layers down, inside the innermost microscopic $D_{1}$ 
\underline{one} proton is about to achieve Planck energy.

However strict Lorentz invariance applies only if space time is uniform 
and flat. Precisely due to gravity $(G_{N} > 0)$ the universe is curved, and 
the horizon grows (since the big bang) only with a finite velocity 
$c < \infty$. 

The idealized nested accelerator and many others which we discuss below are either larger 
than the present universe, are too fragile and breaks down, or are too compact and massive 
and collapses into a black hole. More sophisticated multi-stage devices are flawed by the 
amplification of tiny errors in the process of acceleration or require overly complex
corrective feedback.

Clearly the issue here is not of overall energetics. Just one ton of matter moving at a 
modest velocity of $Km/sec$ has kinetic energy $m_{Pl}c^2 \approx 10^{16} ergs$.
This is also the output in one second of a high intensity laser. The real ``bottleneck''
is the focusing of all of this energy residing initially in $10^{30}$ nucleons or 
photons onto a single proton!

We proceed next to discussion of various type of ``Gedanken accelerators''.

\section {EM Accelerators}

Electromagnetism presents an easily controlled long range force and hence it 
is most natural to consider its utilization for accelerators first. Electromagnetism,
along with quantum mechanics and the mass of the electron, also fix the scale
of atomic (chemical) energies and material strength. Terrestrial EM 
accelerators operating at limited$(R \leq 10 km)$ scales already produce 
$E_{proton} \approx 10^4 GeV$. Similar devices of cosmic dimensions could naively be
envisioned with far larger acceleration energies. Indeed even naturally 
occurring cosmic ray acceleration generates primaries (protons!) with 
$E \approx 10^{12} GeV$ which have already been experimentally observed.

The claim that protons or electrons with energies $E_{Pl} \approx 10^{19} GeV$ can never be achieved is 
therefore far from obvious. Hopefully it will become clearer as we 
proceed through a list of more and more sophisticated accelerators.

\section {Rocket Boosting}

An obvious objection to the notion of maximal energy is its non-Lorentz 
invariant nature. Thus if we put our ``laboratory'' or the accelerator on a rocket and boost
it by a factor $\gamma_R$ the energy that we see in our lab will be enhanced 
by an extra factor of $\gamma_R$ 

Let us assume that the rocket is boosted via emitting ``burnt fuel'' 
at a velocity $\beta$ relative to the rocket. If the latter ever
becomes relativistic then momentum conservation yields (henceforth $c=1$)
\begin{equation}
\label{eq1}
\frac {\Delta m_R \beta} {\gamma_R} = \Delta p = \Delta \gamma_R m_R
\end{equation}
with $\frac {\Delta m_R \beta} {\gamma_R}$ the red shifted 
momentum of the emitted mass element balanced by the increment 
$\Delta \gamma_R m_R$ of the rocket momentum. Eq. (\ref{eq1}) readily yields
\begin{equation}
m_f = m_i e^{- \frac {{\gamma_f}^2} {\beta}}
\end{equation}

Normal exhaust velocities are very small making $m_f$ hopelessly small.
Even if $\beta = 1$, i.e. the rest mass is converted into photons, just to
achieve $\gamma_f \approx 10$ for a 100~kg ``laboratory'' requires 
$m_i \approx 10^{17} m_{sun}$!  Thus no appreciable boost can be achieved and this
naive realization of the ``nested accelerators'' scheme is impossible.

\section{Acceleration via Photon Pressure}
\label{seca}
Instead of accelerating macroscopic pieces of matter, one may 
accelerate single charged particles via photon pressure.
An intense laser beam can accelerate
charged particle to high energies by repeated Compton scattering.  As the
particle approaches the
putative super-planckian regime, it becomes extremely relativistic.  In
the particle's rest frame the photons will be strongly red-shifted (by a
$\gamma^{-1}$ factor) and $\sigma = \sigma_{Thompson} \approx \pi \alpha^2/m^2$
is appropriate.  If the energy flux $\Phi_E$ of the photon beam is:

\begin{equation}
\frac{dW}{dt} = \Phi_E \sigma (1 - \beta ) = \Phi_E \frac{8\pi \alpha^2}{3m^2}
\frac{m^2}{W^2} \approx \Phi_E \frac{\alpha^2}{m^2_{Pl}} \frac{4 \pi}{3}
\end{equation}

The key factor limiting the rate of energy gain is the difference between the 
photon and particle velocities $(1 - \beta ) \approx
\frac{1}{2\gamma^2}=\frac{m^2}{2W^2}$.

Integrating we find that after time $t$,
\begin{equation}
W_f = (4 \pi \alpha^2)^{\frac {1}{3}}(\Phi t)^{\frac {1}{3}}
\end{equation}
The flux of energy in the beam is related to the mean
$\left\langle E^2\right\rangle $ by
\begin{equation}
\frac {1} {4 \pi} E^2 = \Phi.
\end{equation}

We will demand that $E \leq E_{crit} \approx \frac {\pi m_{e}^2} {e}$
so as to avoid vacuum breakdown via $e^+e^-$ pair creation\cite{Sch}.
Hence 
\begin{equation}
\Phi \leq \frac {\pi {m_e}^4}{4e^2} \approx \frac {{m_e}^4}{16 \alpha}
\end{equation}
and
\begin{equation}
W_f \leq (\frac {\pi}{2} \alpha)^{1/3}m_e(m_et)^{1/3} \approx GeV(l/cm)^{1/3}
\end{equation}
where $l \approx t$ is the length of the accelerator. Evidently the growth of energy 
with linear dimension $l, \sim l^{1/3}$ is too slow. Even if 
$l = R_{Hubble} = 10^{28} cm$, $W_f \leq 2.10^9 GeV$. If the lightest charged particle
 were not the electron, but rather some heavier particle x, $W_f$ would be 
larger by a factor of $(\frac {m_x}{m_e})^{4/3}$. 
Planck energies would still be unattainable so long as 
$m_x$ does not exceed $10^5 GeV$!

\section{Circular Accelerators}
\label{ca}
The energy loss due to synchrotron radiation for a circular accelerator is given by \cite{Jac}
\begin{equation}
\label{eq3}
\frac{\delta W}{rotation \: period}= \frac {4 \pi \alpha}{3} \frac {\gamma^4}{R}.
\end{equation}

The mild requirement that $\delta W$ should not exceed $W_{final} = \gamma_f m$ (namely
that all the energy is not radiated away in one turn), implies
\begin{equation}
\label{eq2}
W_f \leq (\frac{4 \pi}{3}\alpha)^{1/3} m (mR)^{1/3}
\end{equation}
or $\gamma_f \leq (mR)^{1/3}$. Eq. (\ref{eq2})is fortuitously similar to Eq. (\ref{eq3}) above
with $m_e$ replaced by $m$, the mass of the accelerating particle. 
Even for $m=m_{proton}$ and $R=R_{Hubble}$\footnote{Again we note that
if we had a stable, massive, larger particle $x$ of mass $m_x \geq
10^6 m_{proton}$ the limit of Eq. (\ref{eq2}) would appear to allow
Planckian acceleration.}

\begin{equation}
W_f \leq 4.10^{13} GeV.
\end{equation}
Another important corollary which follows from the large rate of
synchrotron loss (Eq. (\ref{eq3}) above) is:
Acceleration to an energy W along a circular arc of length L
and an angle $\Theta = L/R$ is impossible unless 
\begin{equation}
\label{eq44}
\theta \leq (\frac {WL}{\alpha})^{1/2} \gamma^{-2} .
\end{equation}
This relation will be useful in subsequent considerations.

\section{Acceleration in 3-D Electromagnetic Fields}

The simplest, ``brute-force'',  accelerator consists of a finite region of space of
dimension R in all directions in which an electric field E exists.  The Final
energies obtained are limited by: 
\begin{equation}
W = eER 
\end{equation}

     The ``3-D'' field in question could be due to two ``capacitor'' plates of size 
R x R placed at a distance R apart.  In principle it appears that $eER \approx m_{Pl}$
can be obtained even for small E if R is sufficiently large.
     A more ``realistic'' setup could be an extreme rotating neutron star such
that the intense B field generates an almost equal E field 
\begin{equation}
\left | \vec E \right | = \left | \vec {\beta} \times \vec B \right | \leq \left |
\vec B \right |
\end{equation}
We will not address the important issues of how one builds cosmic
capacitors or the effects of synchrotron radiation due to magnetic
bending of particle trajectories in the magnetic field around the neutron
star-which drastically reduces the maximal energies obtained.  The interesting
 point is that \underline {regardless of any
 details} no such three dimensional configuration can ,in principle, achieve
  Planck energies.

  The total energy stored in the B and E fields, $\frac{1}{2} [\int E^2 + B^2]
 \approx (B^2 + E^2)R^3$, should not exceed the critical value $Rm_{Pl}{^2}$,
or the whole system will collapse into a black hole of Schwarzchild radius R.  
However, $(B^2 + E^2)R^3 \leq Rm_{Pl}^2 $ implies $ER \leq m_{Pl} $ (or $BR \leq m_{Pl} $) and 
since $e = \sqrt{4 \pi \alpha} \leq 1$, we find that $W_{max} 
= eER $(or  $ eBR) \leq m_{Pl} $, this result is independent of
 how compact and strong or extended and weak the field configuration may be.  The only key 
requirement is that the electromagnetic charge of the elementary particle 
accelerated be smaller than unity.\footnote{The weakly coupled electron, rather than a putative corresponding 
strongly coupled monopole, with $g \simeq e^{-1} >> 1$, should be 
considered as elementary. Then the t'Hooft-Polyakov monopoles are
extended objects of size $\frac{1}{e}\frac{1}{m_M}$ and
effectively contain $\approx \frac{1}{\alpha}$ elementary quanta.}

\section {Linear Accelerators}

The excessive masses and synchrotron radiation can apparently be avoided in 
the case of 
linear accelerators.  Existing LINACs have gains 
\begin{equation}
G \equiv \frac {\Delta W}{\Delta z} \approx e \bar E \approx
10 MeV/meter
\end{equation}
and extend over lengths $L \approx$ few kilometers.  Electron energies $W_e \approx 
50 GeV$ and $\gamma_e \approx 10^5$ can thus be achieved.  Scaling up $L$ 
and/or $G$ one might hope to achieve $W = m_{Pl} \approx 10^{19} GeV $. 

Let us first note that the gain, $G$, is limited by 
\begin{equation}
G \leq G_{max} = eE_{max} \approx \frac {\alpha^3 m_e^2}{2} \approx
Rydberg/Bohr ~radius
\end{equation}

$E_{max}$ is the maximal electric field that can be generated or 
sustained by charges or currents in normal matter.  Stronger fields would 
readily ionize hydrogen atoms. Such fields overcome the work function and skim all 
conduction (valence) electrons breaking all materials.  In general any physical
 system that stores or guides energy with density $U/V = u$ experiences a 
pressure 
\begin{equation}
p \approx const. u
\end{equation}
with a coefficient of order unity.  The maximal E corresponds 
then to the maximal pressure 
\begin{equation}
\label{eqaa}
p_{max} \approx E^2_{max} \approx \alpha^5 m_e^4 \approx 10^{13} 
dynes/cm^2
\end{equation}
that materials can withstand.

Even for $G_{max} \approx eE_{max} \approx eV/A^0 \approx 100 MeV/cm$ the 
minimal, ``net'', acceleration length required for achieving energy $W$ is:
\begin{equation}
L_{min} \equiv W/G_{max} \approx 10 (W/GeV) cm (=10^{20} cm~for~W=m_{Pl})
\end{equation}
Further, to avoid local buckling under the pressure inside the 
pipe the thickness of the pipe, should be comparable with the radius 
$d$.

Let us consider next the electromagnetic field of the accelerating particle
 itself.  The rest frame coulomb field 
\begin{equation}
\overrightarrow{E}^{(0)} = e \hat r / r^2 
\end{equation}
has a coulomb energy
\begin{equation}
W_{coul}^{\left( 0\right) }\approx \int_d^{2d}\vec E^{\left( 0\right)2}
d^3\vec r\approx \alpha /2d
\end{equation}
stored in the pipe wall.  In the lab frame this becomes an 
electromagnetic pulse of duration shortened by Lorentz contraction
down to 
\begin{equation}
\Delta t \approx \frac {d}{\gamma}
\end{equation}
and total energy 
\begin{equation}
\Delta W = \gamma W^{(0)}_{coul} = \gamma \alpha / 2d 
\end{equation}
According to the W.W. method\cite{Jac} this pulse is equivalent to a pulse 
of real photons of energy ($\hbar = 1$) 
\begin{equation}
\omega_{\gamma} \approx \frac {1}{\Delta t} =
\frac {d}{\gamma}  \approx \frac {\gamma 10^{-14} GeV}{d (in~cm)}
\end{equation}
The high energy equivalent photons will produce $e^+e^-$ pairs on 
the wall nuclei with the Bethe-Heitler cross section 
\begin{equation}
\sigma \approx (Z^2 \alpha^3/m_e^2 ) ln(\omega_{\gamma}/m_e)
\end{equation}
This leads to an energy loss at a rate:
\begin{equation}
\frac{\Delta W}{\Delta z} |_{Loss} = \Delta W n \sigma
\end{equation}
It is convenient to parameterize the density of scattering nuclei 
via ($\hbar = 1$) 
\begin{equation}
n = \xi (a_{Bohr})^{-3} = \xi (m_e \alpha)^{-3}
\end{equation}
with the dimensionless $\xi$ being typically $\approx 10^{-2}$.  Demanding that the loss rate does not exceed the 

maximal gain 
\begin{equation}
\frac {\Delta W}{\Delta z} |_{Loss} \leq \frac {\Delta W}{\Delta z}
|_{max gain} = G_{max} \leq \frac {1}{2} m_e^2 \alpha^3 
\end{equation}
yields then using (31) (32) (30) and (28):
\begin{equation}
(\frac{\gamma}{m_ed})\ln(\frac{\gamma}{m_ed}) \leq \frac
{\xi}{\alpha^3 Z^2}
\end{equation}
Thus, for $Z \approx 30$, we find that unless 
\begin{equation}
diameter \approx d \geq 10^8 cm \approx 10^3 km \approx thickness,
\end{equation}
we will not be able to achieve $ W=10^{19} GeV $ for protons.  
This implies in particular that
\begin{equation}
M_{pipe} \approx 2 \rho \pi d^2 L \geq 2.10^4 M_{Sun} (!)
\end{equation}
The large length ($L \geq 10^{20} cm$) and diameter 
($d \geq 10^8 cm$)required exacerbate another crucial difficulty; namely, that of 
guiding the accelerating particle to remain at all times along the central 
axis, and maintaining this linear trajectory with very high precision.  Such
 guidance is required for several reasons. The light-like particle suffers 
a gravitational deflection of order 
\begin{equation}
\delta \theta_0 = gL = \frac {R_{Schw}}{R^2} L
\end{equation}
as it travels a distance $L$ in the field $g$ of a galaxy of Schwarzchild 
radius $R_{Schw}$ at a distance $R$ away.  Using $R \approx 2.10^{24} 
cm$ and $M_{Gal} = 10^{13} M_{Sun}$ as ``typical'' values \footnote {These refer to the 
distance and mass of the Andromeda galaxy.} and
 $R_{Schw} = G_N M_{Gal} = 10^{13} R_{Schw. Sun} = 2.10^{18} cm$ we find by comparing with
 Eq. (\ref{eq3}) that the radiative synchrotron losses, due to the gravitational bending
 will be excessive once $L \geq 10^{19} cm$, one order of magnitude smaller
 than $L_{min} \approx 10^{20} cm$.  

More generally, let us envision some 
transverse ``wobbling'' of the accelerated particle around the axis with 
wavelength $\lambda$ and amplitude $\delta$. Using Eq. (\ref{eq44})
(with $\Theta \sim 
\frac{\delta}{\lambda}, L \sim \lambda, W=\gamma
m_N, \gamma \approx 10^{19}$) we find 
that in order to avoid excessive synchrotron losses we must satisfy
\begin{equation}
\delta \leq 3\lambda (\lambda m_N)^{\frac{1}{2}} 10^{-28}.
\end{equation}
  Even if we take $\lambda \approx d = 10^8 cm$
, an incredible accuracy of trajectory with 
\begin{equation}
\delta \leq 10^{-9} cm
\end{equation}
should be maintained at all times.  This would seem to be 
virtually impossible if we indeed maintain an empty pipe hole and guide 
the particle only via fields generated at the pipe wall at a distance of $10^8 cm$ away.

A key observation in this context is that on-line monitoring and 
correction of the orbit is impossible.  Thus let us assume that at some ``station'' $S_N$ we
 find the position and velocity of the accelerating particle and then 
transmit the information to the next ``station'' $S_{N+1}$ ``downstream'' so 
that some corrective action can be taken there.  If the distance between 
the two stations is $\Delta L$ then a signal traveling with the 
velocity of light will arrive only a very short time 
\begin{equation}
\delta t = \frac {\Delta L}{c} (1 - \beta) = \frac {\Delta L}{2 c \gamma^2}
\end{equation}
ahead of our particle. The distance that a corrective device 
can move at that time is only
\begin{equation}
\delta = \frac {\Delta L \beta_D}{2 \gamma^2}
\end{equation}
where $\beta_D$ is the velocity with which the corrective device 
moves.

Using $\gamma = 10^{19}$ we find (even for $\beta_D = 1$ !) that
\begin{equation}
\delta \leq \frac {\Delta L}{10^{38}}
\end{equation}
so that even for $\Delta L \approx L \approx  10^{20} cm, \delta \leq 10^{-18} cm$!

Our considerations of the linear accelerator touched only some of the 
possible difficulties. Yet these 
arguments strongly suggest that such a project is not merely difficult 
because of, say, the need to assemble a $10^{20} cm$ pipe of 20,000 solar masses. 
 Rather there are inherent, ``in principle'', difficulties  which make 
the project of accelerating particles to Planck energy via a linear 
accelerator impossible.

\section  {Gravitational acceleration}

Gravity is, in many ways, the strongest rather than the weakest
interaction. This is amply manifest in the gravitational collapse to a 
black
hole which no other interaction can stop. Along with the very definition of
 the Planck mass, $ m_{Pl}$, this naturally leads us to consider gravitational accelerators, and
the acceleration (or other effects) of black holes in particular.

\section{Direct Acceleration}
\label{da}

  If a particle of mass $\mu$ falls from infinity to a distance r from
the center of a spherical object of mass m, it
 obtains, in the relativistic case as well, a final velocity
\begin{equation}
\label{eq4}
\beta_f=\sqrt{r_{SW}\over r}
\end{equation}
  with $r_{SW}={G_N m\over c^2}$, the Schwarzchild radius of the mass m.
In order that our particle obtain, in a ``single shot'', planckian
energies, we need that $\epsilon\equiv1-\beta_f=\frac{1}{2\gamma_f^2}=\frac
{\mu^2}{2m_{Pl}^2}$ or $\epsilon=\frac{r-r_{SW}}{r_{SW}}
=\frac{\mu^2}{2m_{Pl}^2}.$  The last equation applies also if at infinity
we have initially
 a photon or massless neutrino of energy$\mu$.
Using for the generic starting energy or rest mass $\mu \leq m_N \approx 1 GeV$ we
then find
\begin{equation}
\label{eqac}
\epsilon \equiv \frac {r - r_{SW}} {r_{SW}} \leq 10^{-38}
\end{equation}
since to avoid trapping the last ratio should exceed 2 this Planck acceleration is clearly
ruled out.

\section {Hanging Laboratory}

The difficulty of using the gravitational fields in the 
neighborhood of the black hole horizon is best illustrated by considering 
the following concept of a ``hanging laboratory'' (suggested to us by N.
 Itzhaki as a possible counterexample to the impossibility of achieving 
Planck energy).  Thus let us envision a black hole of very large radius 
$R_{Schw}$.  We can then hang our laboratory keeping it at a distance d 
away from the horizon.  If $R$ is sufficiently large there will be minimal 
tidal distortion across the laboratory.  An energetic particle falling 
towards our laboratory would then be blue-shifted by a factor $R/d$ and can
 appear superplanckian to us, once $\frac {R}{d} $ is large enough.

Amusingly, this fails simply because any ``rope'' used to hang the laboratory
 and extending between $aR_{Schw}$ and $bR_{Schw}$ with $a \geq b 
\approx O(1)$ tears under its own weight.  Imagine that the rope 
is ``virtually displaced'' downwards around the upper hanging area by some 
distance $\delta z$ which we take to be $a_{0}$, the inter-atomic 
distance in the rope material.  This will decrease the gravitational energy by
\begin{equation}
\label{eqad}
\delta m  (\frac {1}{a} - \frac {1}{b}) \approx O(\delta m)
\end{equation}
with $\delta m$ the mass of a rope element of length $\delta z$ :
\begin{equation}
\delta m = S \rho \delta z = S(M(A,Z)/a{_0}{^3})a{_0}
\end{equation}
where S is the cross-sectional area of the rope and 
$\rho = M(A,Z)/a{_0}{^3}$ the mass density with $M(A,Z) = Am{_N}$ the 
average mass of the nuclei in the rope material.  A displacement by 
$\delta_z=a_0$ causes the tearing 
of the rope.  The energy required for tearing the rope is  $N \epsilon$ with 
\begin{equation}
N \approx S/a{_0}{^2}
\end{equation}
the number of atomic nearest-neighbor bonds torn along the z direction and $\epsilon 
\approx \frac {1}{2} m_e \alpha^2 $ is our usual estimate for the 
bond energy.  Tearing will be avoided if the gravitational energy
gained (which from Eq. (\ref{eqad}) is 
$\approx \delta m$) cannot supply the $N \epsilon$ energy required i.e. if:
\begin{equation}
N \epsilon \geq \delta m \approx NM(A,Z) \approx NAm_N,
\end{equation}
which leads to the condition (note that as $a_0$ cancels):
\begin{equation}
\frac {\alpha^2 m_e}{2 m_N} = 1.4.10^{-8} >  A \geq 1
\end{equation}
This inequality clearly fails.  
We can also show that tearing cannot be avoided by having the
 cross section of the rope change so that a broader section on top can more readily 
sustain the weight of a lighter section on the bottom.

\section{The Unruh Accelerator}
\label{tua}

A sophisticated accelerator using repeated ``slingshot kicks'' was suggested
 by Unruh. This beautiful concept is best
illustrated in the following
simple two black holes context. Consider first two black holes of equal 
mass
$m_1=m_2=m$  at points $P_1 $ and $P_2$ located at $ +L,-L$ along the
$z$ axis.  A relativistic neutral particle $\mu$ (neutron, neutrino,or photon) 
is
injected parallel to the z-axis with some relative impact parameter near $z=0$.
With an appropriate choice of the impact parameter $b=b_0$, the accelerated
particle describes a ``semi-circle'' trajectory around $m_1$ at $P_1$,
and is reflected around this mass by an angle $\theta=\pi$ exactly.
Moving then along a reflected ($ x\to -x$) trajectory the particle $\mu$
approaches the other mass $m_2$ at $P_2$, and is reflected there by
$\theta=\pi$ as well. The particle will eventually describe a closed geodesic
trajectory bound to the two mass $ m_1, m_2$ system.  In reality these two
masses move. For simplicity consider the case when the masses move symmetrically towards
each other with relative velocity $\beta$. Transforming from the rest mass of $m_2$
say to the ``Lab frame'', we find that in each reflection the energy of $\mu$ is
enhanced according to $W_\mu \to  W_\mu\sqrt{(1+\beta)\over (1-\beta)}$.  The
last equation represents the boost due to the ``slingshot kick'' alluded to above.
If we have $N$ such reflections, the total boost factor
is $\sim(\frac{1+\beta}{1-\beta})^{\frac{N}{2}}$.  In order to achieve
Planck energies, for starting energies $\sim GeV$, this overall boost
should exceed $10^{19}$.  
However, in this simple geometry the
total number of reflections is limited by $N=1/\beta$.
After more reflections, the two masses will either coalesce or reverse their
velocities, leading now to a deceleration of the particle $\mu$  upon each
reflection. Furthermore, in order to avoid the two masses coalescing
into a black hole upon first passage, we find that $\beta_{max}$, the
maximal $\beta$ along the trajectory, is $\beta_{max} =
\frac{2}{3\sqrt{3}}$. The total amplification is therefore bound by 
$(\frac{1+\beta_{max}}{1-\beta_{max}})^{\frac{1}{2\beta_{max}}}\leq
2.6$.

The Unruh set up involves,
however, two additional heavier black  holes $M_1=M_2=M $with $m_1,m_2$,
revolving around $M_1,M_2$ respectively, in circular orbits of equal radius
$R$ and period $T=2\pi R/\beta$, with $\beta $ the orbital
velocity. The two orbits are assumed to lie in the $ (x-z)$ plane with the
centers of the circles located at $(x,z)=(0,+L)$. The ``top points''
on the two circles, {\it i.e.} the points where x is maximal, define now
the original reflection centers  $ P_1,   \ \ P_2 = (R,-L),\ \  (R,+L) $.
The oppositely rotating masses $m_1$ and $m_2 $ are synchronized to pass at
$P_1$ and $P_2$, respectively, at the same time -- once during each period T.
Furthermore, the motion of the accelerated mass $\mu$
is timed so as to have $\mu$ at the extreme left point on its ``Stadium
Shaped'' orbit $(x,z)=(R,-L-\rho)$ or, at the extreme right point
$(x,z)=(R,L+\rho)$, at precisely the above times.  This then allows us to
achieve the desired sling shot boosts, repeating once every period T. Note that
2T is now the period of the motion of the overall five body system $(M_1,M_2,m_1,m_2
;\mu)$.

However, the inherent instability of this motion limits the
number $N$ of periods (and of sling shot boosts)
and foils this ingenious device.
The assumed hierarchical set-up $ L>> R>>r_{SW}\approx b_0$ can be used to approximate
the angular deflection of $\mu$ while it is circulating around $ m_1 $, say, by
\begin{equation}
\theta = \int_{u_{min}\simeq0}^{u_{max}} {du \over
\sqrt{ {1\over b^2} - u^2( 1-2G_N m u) }}
\end{equation}
with $b$ the impact parameter and $u_{max}=1/\rho$ corresponding to the
turning point of closest approach. Independently of the exact (inverse elliptic
function) dependence of $ \theta$ on $b/{r_{SW}},\ \rho/{r_{SW}}$, a fluctuation
 $\delta b^0$ around the optimal $ b^0$, for which $\theta$
equals $\pi$, causes a corresponding fluctuation in the reflection angle 
$\theta$ : $ \delta^1\theta
=\pi-\theta=k\delta b^0/b^0$, with the dimensionless constant $c$ of
order one. The large distance $L$ transforms this small $\delta\theta$ into
a new impact parameter deviation, $\delta^1(b) = L\delta^1\theta$. The ratio
between successive deviations of the impact parameter is then given by
 $|\delta^{1}(b)|=(cL/b_0)
\delta^0(b)$, etc. After $N$ reflections, we have therefore
\begin{equation}
\label{eqcc}
\delta^N(b) \simeq (kL/b_0)^N \delta^{(0)}(b)
\end{equation}
For $ L>R>b_0$, the ratio in the last equation $cL/b_0>>1$. 

The individual sling-shot gain $\sqrt{\frac{1+\beta}{1-\beta}}$
depends on the linear velocity $\beta$ of the circular motion of $m_1$
around $M_1 = M$ (or $m_2$ around $M_2=M$).  Let $R_{SW}$ be the
Schwarzchild radius of the large mass $M$.  For circular motion with
radius $R$ the above $\beta$ is given by
\begin{equation}
\beta=[\frac{R_{SW}}{2(R-R_{SW}}]^{\frac{1}{2}}
\end{equation}
In order to avoid capturing $m_{(i)}$ on $M_{(i)}$ we need that $R\geq
\frac{3}{2}R_{SW}$ i.e. $\beta \leq \beta^{circ}_{max} \leq
\frac{1}{\sqrt{2}}$.  Since we like the motion of $\mu$ to be
dominated by $m_i$ and avoid its ($\mu$'s) falling towards the large
masses $\frac{R_{SW}}{R}$ and $\beta_{max}$ have to be much larger.
However even with the above $\beta_{max}$ we need $N\approx54$
repeated sling shot kicks to achieve an overall
$[(1+\beta_{max})/(1-\beta_{max})]^{\frac{N}{2}} \approx 10^{19}$ enhancement.
To avoid
complete orbit deterioration for the accelerating particle $\mu$, {\it i.e.}
to avoid $\delta^N(b)\simeq b_0$, we need then, according to Eq. (\ref{eqcc}), even for a modest 
$(cL/b_0) \approx 10$, an initial precision
$\delta b^0/{b_0}=10^{-50}$, which, in particular, is exceeded by the quantum
uncertainty in $b$. (This beautiful refutation of the Unruh accelerator is 
due to B. Reznik.)

The above Unruh accelerator is actually a prototype of many other 
gravitational accelerators in which the high energy is achieved by 
repeating many stages of more limited boosts.  Another example, 
suggested to us by A. Polyakov, involves the 
phenomenon of super-radiance\cite{Zeld}.  Specifically if two particles fall towards
 a rotating black hole, and collide in its 
vicinity then one can be emitted with an energy higher than the total energy. 
It turns out again that the ratio of $E_{final}/E_{initial} 
\leq 1 + \epsilon$ with $\epsilon \approx 0,2$  so 
that many repeated such collisions with a series of Kerr black holes is 
required and again exponentially growing fluctuations are encountered.
  
Thus, ab-initio, finely tuned, perfect, gravitational accelerators are impossible. 
Could we still achieve such energies if we monitor the trajectory and correct deviations?
Since the accelerated particle is neutral (to avoid synchrotron losses), the correction of
the trajectory requires additional mass(es).  Just adding such masses only complicates the 
problem turning the Unruh accelerator into say, a six (or more) body problem.  Hence we
need two ingredients:
(i) extra non-gravitational forces to navigate the corrective masses and
(ii) means for monitoring the trajectory of the accelerated particle.
Neither of these tasks seems readily achievable.  Thus to have an appreciable light bending
effect the corrective mass should be large - probably a black hole itself and its propulsion
via non-gravitational forces (say some rocket mechanism) appears impossible

The second task of monitoring the location and momentum $\overrightarrow {r} (t),\hat p (t)$
of the accelerated particle $\mu$ is even more difficult.  To this end we need to scatter
other particles from $\mu$.  Since, among others, we need to correct for the effect of 
$\underline {quantum fluctuations}$ we have to monitor each accelerated particle
individually, by having several scatterings from $\mu$ during one traversal of its closed
orbit.

Such scatterings, being intrinsically quantum mechanical, introduce further, uncontrolled,
perturbations.  Furthermore, these scatterings systematically deplete the energy of the 
accelerated particle $\mu$.  For concreteness take $\mu = \gamma$ (photon) and assume we
scatter electrons from it. In any $e \gamma$ collision, be it elastic or inelastic, the
initial photon retains only a $\underline {fraction}$ of its energy.\footnote {This follows
essentially from kinematics: when the energetic photon scatters ``elastically'' on an
electron at rest it retains its direction in the Lab frame.  However if in the center of
mass the scattering is by an angle $\theta^*$, then the ratio of the final and initial photon 
energies, in the Lab, is $\frac {W'}{W} \approx \frac
{1+cos\theta^*}{2}$.  Since in the center of mass
frame the Klein-Nishima formula for the Compton process yields roughly an isotropic
distribution we have $\langle \frac {W'}{W} \rangle \approx \frac
{1}{2}$.  

In passing we
note that if the particles with which the accelerating particle $\mu$ collides have similar
energy and (for $m_{\mu} \neq 0$) similar mass then the elastic collision yields in the
final state a more energetic and a slower particle.}

\section{Evaporating Black Holes}
\label{sec12}

Hawking evaporation of black holes naturally leads 
to single quanta of energies approaching (but not exceeding!) $m_{Pl}$. Hawking found\cite{Haw}
that a black hole has an effective temperature $T_{BH} \approx \frac {1} {R_{BH}} \approx
\frac {{m_{Pl}}^2} {m_{BH}}$ with $R_{BH}$, $m_{BH}$ the radius and mass of the black hole.
As $R_{BH}$ approaches $l_{Pl}$, the temperature $T_{BH}$ approaches $m_{Pl}$, and 
photons, or other quanta, with energies~$\approx T_{BH} \approx m_{Pl}$ could, in 
principle, be emitted. However, precisely at this point, also the total mass of the 
black hole approaches $m_{Pl}$ and energy conservation forbids the emission of 
several such quanta or one quantum with $W>>m_{Pl}$.

If as it emits the last quanta, the center of
mass of the black hole had an appreciable boost, say $\gamma \geq 3$, then the quanta 
emitted in the direction of motion of the black hole could be Doppler shifted and have
superplanckian energies: $W' = \gamma W_{in black hole frame} \geq m_{Pl}$.
The recoil momentum accumulated through
the Hawking radiation is, at \underline{all} stages, $P_{Rec}
\approx m_{Pl}$\footnote{This amusing result is very simply explained. Consider the
overall recoil momentum accumulated when the black hole loses half its initial
mass via $N \approx \frac{m_{BH}}{T_{BH}}\approx (\frac{m_{BH}}{m_{Pl}})^2$ quanta.
The ``random'' vectorial addition of the $N$ recoil momenta yields $\left | \vec P_{Rec}
\right | = \left | \vec K_1 + \ldots \vec K_N \right | \approx \sqrt{N}K = 
\sqrt{N}T_{BH}\approx m_{Pl}$.}, so that $\gamma_{Rec} \sim 1$, and
no appreciable extra boost effect is expected.  Note that $P_{Rec} \approx m_{Pl}$ implies
a recoil kinetic energy of the black holes $W_{recoil} \approx \frac{P^2}
{2M_{BH}} = \frac{m{_{Pl}}{^2}}{2M_{BH}} = T_{BH}$ as required by equipartition.

In principle, the black hole can be directly boosted so that the extra Doppler shift yields
transplanckian energies for the last photons.  In reality this boost is rather
difficult to achieve.  If the black hole is to evaporate in Hubble time its
initial mass and radius cannot exceed $m^{(0)}_{BH} \approx 10^{15} g $ and
$R^{(0)}_{BH} \approx 10^{-13} cm$ respectively.  Thus even a single electric
charge on the black hole creates a field $eE \approx \frac {\alpha}{R_{BH}^2}$
which exceeds the vacuum breakdown limit  $eE \approx m_e^2 $ by a factor of
about a hundred already for the initial radius.  Hence the black hole will
immediately lose its charge via vacuum $e^+e^-$ pair creation and electromagnetic 
acceleration is impossible.

  Finally, gravitational boosting of the mini black holes is excluded by the reasoning
presented in Sections (\ref{da}),(\ref{tua}). In
particular since the minimal distance to which the mini black hole can approach
the big black hole without being captured onto it, is limited by $r_{Schw}$, the Schwarzchild radius of the
large black hole, we cannot achieve even $\gamma = 2$ boosts.

\section {Can Small Black Holes Be Created?}
Another serious difficulty with using evaporating mini
black holes as Planck accelerators - albeit for O(1) quanta emitted at the
last stage - stems from the fact that such black holes cannot be created 
ab-initio in the lab.  The bottleneck is again the need to focus excessive
energy onto a tiny domain of size $R_{Schw}$ - the Schwarzchild radius of the
prospective black hole.
  
To see that, consider the following ``attempt'' to build small black holes by focusing
energy.  Imagine a spherical arrangement of $N$ high
intensity lasers each of cross section $\approx d^2 $ and wavelength
$\lambda$  on a large spherical
shell of radius $L$. It is designed to 
focus the energy emitted by the lasers which are all directed radially inward 
into a hot internal spot of diameter $d$.
The number of lasers $N$ is restricted by
\begin{equation}
N \leq 4 \pi (\frac {L}{d})^2
\end{equation}
Let the energy density in each laser beam be $\rho_0 \approx E^2/2$. We can achieve
$N$ -fold enhancement of this density in the central spot of radius $d$
\begin{equation}
\rho_{center} \approx N\rho_0 = NE^2/2
\end{equation}
This generates a black hole of radius $d$ provided that
\begin{equation}
M_{(inside~d)} = \frac {4 \pi d^3 \rho_{center}}{3} = \frac {2 \pi}{3} NE^2 d^3 \geq
2 m_{Pl}^2 d 
\end{equation}
or
\begin{equation}
\label{eq5}
\sqrt{N} E d \geq \sqrt {\frac {3}{ \pi}} m_{Pl}.
\end{equation}
Diffraction limits the size $\Delta$ of the spot image of any of the individual
laser apertures according to:
\begin{equation}
\Delta \geq \frac {L}{d} \lambda \approx \sqrt {\frac {N}{4 \pi}} \lambda.
\end{equation}
Demanding that $\Delta$ not exceed the original aperture $d$
which is also the assumed black hole radius implies then
\begin{equation}
\label{eqae}
N \leq  4 \pi d^2 / \lambda^2 
\end{equation}
The upper bound on $E$, required to avoid vacuum breakdown 
\begin{equation}
E \leq m_e^2 / \sqrt{\alpha} ~(or~m_e^2 \alpha^{\frac {5}{2}})
\end{equation}
and Eqs. (\ref{eq5}) and (\ref{eqae}) imply
\begin{equation}
\sqrt {\frac {3}{\pi}}m_{Pl} \leq \sqrt{N} Ed \leq
\sqrt{4 \pi} \frac {d^2}{\lambda^2} \frac {m_e^2}{\sqrt{\alpha}}
\end{equation}
or
\begin{equation}
\label{eqde}
d \geq \sqrt {\frac {3}{\pi}} (\frac {m_{Pl}}{m_e})^{\frac {1}{2}}
\alpha^{\frac {1}{4}} \sqrt {\frac {\lambda}{m_e}} \approx 10^{10} \sqrt {\frac 
{\lambda}{m_e}}
\end{equation}
demanding that the evaporation time of our black hole be shorter than
the Hubble time implies
\begin{equation}
d^3 m^2_{Pl} \leq R_{Hubble} \approx 10^{28} cm.
\end{equation}
If we use this along with 
\begin{equation}
d^3 m^2_{Pl} \geq 10^{30} \frac{m^2_{Pl}}{w^{\frac{3}{2}}_x
m^{\frac{3}{2}}_e}
\end{equation}
(with $w_x=\frac{1}{\lambda_x}$ the photon's energy) which readily
follows from Eq.(\ref{eqde}) we conclude that 
\begin{equation}
W_{\gamma}\geq 10^{20} GeV
\end{equation}
That is we need super-Planckian photons to start with!
There is the ``standard'' mechanism of generating black holes via the
collapse of supermassive ($m_{core}\geq
(2-3)m_{Chandrasekhar}\approx m_{BH}$) stellar cores with
$m_{Chandrasekhar}\approx \frac{m^3_{Pl}}{m^2_N} \approx 1.4 M_{\odot}$, the
Chandrasekhar mass.  The evaporation time of such a black hole is:
\begin{equation}
t_{evap}\approx (\frac{m_{Pl}}{m_N})^6 t_{Pl} \approx 10^{70} sec
(\approx 10^{53}t_{Hubble}).
\end{equation}
It has been speculated by Lee Smolin [\cite{Smolin}] that the final
decays of these black holes will spawn new universes.

\section{The ``Inverted Cascade Accelerator'' Utilizing Repeated Particle Collisions}
Most of the limitations and bounds on accelerators that were found above stem
from the fact that our starting particles have some limited energy \footnote{
In this section and the next we will use $E_0, E_k,$ etc. to denote the energy of
individual particles and $W_0, W_k,$ the energy of the complete system with $N$ particles},$E_0$
 and that normal materials have limited energy density and strength
$\approx E_0^4 $.  We may try to generate an ``inverted cascade'' process where
fewer and fewer particles will survive at consecutive stages but with increasing
energy.  
\begin{equation}
E_{k} = \lambda_k E_{k-1};~~~\lambda_e \geq 1 
\end{equation}
Since the starting energy in each stage is higher, the original
energy-strength bounds do not apply. This is a crucial difference between this Gedanken 
accelerator and the, superficially similar, rocket accelerator.  There even later stages are 
constructed from ordinary fragile material.  Indeed we do not envision here a 
static accelerator, but rather a sequence of transient or ``single-shot'' devices.  It is
modeled in an abstract way after the schematic ``nested accelerator'' of Fig. (1).
We are, however, building consecutively the various stages of the accelerator
starting from the biggest outermost stage first as the very acceleration process
proceeds.  In this ``dynamical accelerator'' earlier stages which have fulfilled their
mission literally ``evaporate'' and disappear.

Let us assume that the process starts at the zeroth stage with $N_0 $ particles,
 each with kinetic energy $E_0 $, so that the total initial energy is
\begin{equation}
W_0 = N_0 E_0 
\end{equation}
In the kth stage we have $N_k $ particles of average energy $E_k $,
and the total energy at this stage is
\begin{equation}
W_k = N_k E_k 
\end{equation}
If no ``waste heat energy'' is dissipated in going from one stage to the
next then $W_k = W_0 $ and after $K$ stages all the energy 
would concentrate
in one single particle of energy
\begin{equation}
E_{K} = \prod^K \lambda_{k} E_0 = N_0 E_0 = W_0 
\end{equation}
However such a process - where all the many, low energy final,
particles conspire to reconstitute the energetic primary proton is completely
impossible.  Specifically, we need to worry about the second law of thermodynamics
as well.

Thus along with a fraction $N_{k+1}/N_{k} $ of the particles of the kth 
generation which have been elevated to higher energies $E_k \rightarrow 
E_{k+1} = \lambda E_k $ and constitute the (k+1)th stage of the accelerator,
we need to emit a certain minimal amount of energy as ``waste
heat'' in the form of particles of energy lower than $E_K$. What is  
the maximal efficiency of the kth stage accelerator, i.e. what is
the maximal value of the ratio  
\begin{equation}
\epsilon_k \equiv \frac{W_{k+1}}{W_k} = \frac {N_{k+1} E_{k+1}}{N_k E_k}?
\end{equation}
Let us identify the average
energy of a particle in the kth stage with a fictitious `` effective temperature'' $T_k $
for this stage.
\begin{equation}
T_k = E_k.
\end{equation}
Let us follow the pattern of our schematic idealized nested accelerator
of Fig. (1).  Thus we identify the kth accelerating stage as a device which
transfers a fraction $\epsilon$ of its energy  to the next,
$k+1^{th}$ stage and at the same time dissipates the rest.
The maximal thermodynamic efficiency of such a device
 is limited by:
\begin{equation}
\label{eq63}
\epsilon_k \leq (2-T_k/T_{k+1})^{-1}
\end{equation}
or
\begin{equation}
\epsilon_{k-1} \leq \frac{T_k}{T_{k+1}} (1-\frac{T_{k+1}-2
T_k+T_{k-1}}{T_{k+1}})^{-1}
\end{equation}
Hence the equation is bound by
\begin{equation}
\label{eqff}
\epsilon \equiv \Pi \epsilon_k \leq \frac{T_0}{T_k} \prod^K
(1-\frac{T_k''}{T_{k+1}})^{-1}\approx \frac{T_0}{T_k}=\frac{E_0}{E_{max}}
\end{equation}
where $T''_k$ indicates a second derivative with respect to $k$ and we
 assumed that we have many stages
 with $\delta T_k \equiv
T_{k+1}-T_k \ll  T_k$

In Eq. (\ref{eqff})  $E_{max} \approx E_k$ is the final, maximal, energy achieved.
To achieve $E_{max} $ starting with particles of energy $E_0 $, we need
to have initially at least $((E_{max}/E_0)/\epsilon )$ particles i.e. $N_0 \geq
(E_{max}/E_0)^2 $.  This result is independent of the amplification
$\lambda $ and the corresponding total number of stages $K$ required  
so long as $K$ is sufficiently large and the enhancement ratios
$\lambda_k$ sufficiently close to one so as justify the approximation
used in Eq. (\ref{eqff}).

We have not succeeded in constructing (even Gedanken!) mechanical or EM inverted cascade 
accelerators with the above efficiency.  Roughly speaking the energetic particles in the
$k^{th}$ stage tend to disperse transversally. A long range, coherent, attractive force
seems to be required in order to keep these particles confined.  Gravity can precisely supply
that.  Indeed the evaporating black hole of section 9 can be viewed as an ``inverted cascade''
Planck accelerator.  To this end we should view its initial stage with mass 
$W_0 = M^{(0)}_{BH} = R^{(0)}_{BH} m_{P}^2$ and corresponding temperature $T^{(0)}_{BH} \approx
\frac {1}{R^{(0)}_{BH} } $ as a collection of $N^{(0)}$ Hawking photons each with average
energy $T^{(0)}_{BH} $ so that all together we have
\begin{equation}
\label{eq66}
N_0 = \frac {W_0}{T_0} \approx (R_{BH} m_{Pl})^2 \approx (\frac {R_{BH}}{l_{Pl}})^2 
= (\frac {m_{Pl}}{T_0})^2 
\end{equation}
In the process of Hawking radiation some of these ``photons'' are radiated away - the total
energy decreases, but the remaining, fewer, photons get ``hotter'' i.e. more energetic
according to: $E \approx T \approx \frac {1}{R_{BH}} \approx \frac {m_{Pl}^2}{W}$ until
ultimately we stay with $O(1)$ Planck photons $W_{final} = E_{final} \approx m_{Pl}$, and
Eq. (\ref{eq66}) precisely conforms to the above $N_0 =
(E_{max}/E_0)^2$ with $E_{max}=m_{Pl}$.\footnote{\label{foa}The optimality of the
black hole in ``upgrading'' energy up to a Planckian level may be
related to another interesting issue of maximal number of elementary
``computations'' in a given spacetime region. (S. Massar and S. Popescu, work in preparation.)}

\section{A Conjecture on Maximal Energy of Accelerators}
The accelerators considered so far - with the exception of the
 evaporating black hole - 
 fall into two basic categories:
accelerators where a large number of quanta of some common low energy $E^{0} $
are absorbed by the accelerating particle, and those involving predesigned,
classical field configurations.  The latter include neutron stars, the Gedanken 
capacitor fields and also the linear accelerator.
Much energy is stored in advance in coherent collective classical degrees of
freedom.  These degrees of freedom then directly interact with and accelerate
the particle.  

        The photon beam accelerator of section \ref{seca} is typical of
the first category.  For all accelerators in the first category there is a
clear-cut limit on the rate of energy increase. It follows from the confluence
of the limited energy density or energy flux $\Phi \leq E_0^4 $ and the
$(1-\beta) \approx \frac {1}{2 \gamma ^2} $ relative velocity factor.  If
$E_0 $ is some effective low energy physics mass scale ( = energy of photons,
atomic energies, nuclear energies, nucleons mass,...) then this implies
\begin{equation}
\frac {d \gamma}{dt} \leq \frac {E_0}{2 \gamma^2}
\end{equation}
and $t_{final} $ the time required to achieve $\gamma_{final} $ is therefore
\begin{equation}
\label{eqb}
t_f \geq \frac {\gamma_f^3}{E_0}
\end{equation}
The issue of acceleration via coherent preexisting fields is less clear.
The energy density in the accelerating field is still limited by $u \leq
E_0^4 $ with $E_0 $ some low energy physics scale.  In addition to this we also need to address the
questions of how the classical field is to be generated in the first place and
also of the stability of the trajectory of the accelerating particle.  We would like to speculate that given
all these limitations the general bound Eq. (\ref{eqb}) still applies.  

It is amusing to apply this speculation to super-high cosmic
rays. Recently such primary cosmic rays, with
energies exceeding $3\cdot10^{11}$GeV have been
observed\cite{Linsley}\cite{Bird}. These findings appear to conflict
with the Greisen-Zatsepin-Kuzmin bound on \underline{cosmologically}
originating primary protons (because of the energy degradation by the
$3^{\circ}$ background radiation). However Eq. (\ref{eq66}) (with a
``Natural'' choice $E_o\simeq m_N\simeq GeV$) would allow, in Hubble
time, $\gamma_f$ values up to
\begin{equation}
\gamma^{max}_f = (R_{Hubble} \cdot GeV)^\frac{1}{3} = 10^{14}
\end{equation}
still exceeding the maximal value observed by $\approx 100$.

Needless to say the maximal cosmic ray energy observed to-date need
not be the true absolute end of the cosmic ray spectrum. This raises
an interesting issue. What if protons (or neutrons) of energy $\geq
10^{14}$ GeV are ever found and our speculated upper upper bound value
(Eq. (\ref{eq66})) indeed applies? We would then be forced to the
radical conclusion that these protons/neutrons must originate from the
decays of long lived particles x of mass $m_x\geq 10^{15}$GeV.

\section{Hubble Time and Mass Dependence of the Maximal Energy Achievable}

Certain Planck accelerators have been ruled out by the fact that the Hubble radius of the 
universe is too small to accommodate them.  Present observations and estimates of 
$\Omega(\equiv \frac {\rho_{cosmic}}{\rho_{critical}})$ tend to favor an open, or critical,
universe which keeps expanding forever.  It would seem therefore that after waiting a 
sufficiently long time ($t \approx 10^{15} t_H $ according to the
scaling law of Eq. (\ref{eqb}) with
$E_0 \approx 1 GeV$ acceleration to super-Planckian energies becomes feasible.

        We observe, however, that truly cosmic accelerators are
``red-shifted'', in the process
of the expansion.  This red shift reduces the rate of energy increase $\frac {dW}{dt}$ in such a
way that even waiting for an infinite time will not enhance the final energy obtained by more 
than a factor of order one.

        To see this, consider, as an example, the photon beam accelerator of Section (4).  Here
$\frac {dW}{dt} \approx \frac {\Phi \alpha^2}{W^2}$ decreases with the expansion, simply
because $\Phi$, the energy flux in the beam red-shifts according to

\begin{equation}
\Phi(t) = \Phi(t_H)[a(t_H)/a(t)]^4
\end{equation}

with $a(t)$ the scale factor and $t_H \approx 10^{19} sec$, the ``present'' time.  Using
$a(t)=a(t_H) (t/t_H)^{\frac {2}{3}}$ - appropriate for a matter dominated universe - 
we then have

\begin{equation}
\frac {dW}{dt} = \frac {\Phi(t_H) \alpha^2}{W^2} (\frac {t_H}{t})^{\frac {8}{3}}
\end{equation}

which integrates to 

\begin{equation}
W^3(t) - W^3(t_H) = \frac {9}{5} \Phi_0 \alpha^2 t_H [1-(\frac {t_H}{t})^{\frac {5}{3}}]
\end{equation}

Thus, waiting for $t=\infty$ rather that for $t=2t_H$ will enhance the final energy gathered
in the waiting period only by $[1-(\frac {1}{2})^{\frac {5}{3}}]^{-1}$ i.e. by 
$\leq 50\%$.  In passing we note that since at present $a(t) \approx t$ implying that in
a few \( t_H \)'s the universe will be curvature dominated and $a(t)$ starts growing at a faster,
linear, rate.  Also considering Eq. (\ref{eqb}) we note that even drastic reduction of the 
red-shifting of $\frac {dW}{dt}$ to $\frac {dW}{dt} = \frac {\Phi(t_H) \alpha^2}{W^2} 
(\frac {t_H}{t})^p$ with $p \geq 1$ still will allow only a $(\ln \frac {t}{t_H})^{\frac {1}{3}}$
increase of $W_{final}$.

        Our statement that the horizon is too small really means that the dimensionless Dirac
number: $Di = R_{Hubble}m_e(m_N) \approx 10^{37} - 10^{40}$ is in some sense ``small''.  In
particular it does not much exceed or is smaller than other dimensionless combinations that
naturally arose in our previous discussions such as $N_0 = (\frac {m_{Pl}}{E_0})^2 \approx
(\frac {m_{Pl}}{m_e})^2 or (\frac {m_{Pl}}{m_N})^2 = 10^{38} - 10^{40}$, the minimal number of
particles in an ideal inverted cascade accelerator or $N_0' = (\frac {m_{Pl}}{E_0})^3 \approx
10^{57} - 10^{66}$ which arises in the discussion of section 12.  If $m_{Pl}$ is left fixed,
which we assume to be the case, we can enhance the Dirac combination relative to $N_0, N_0'$
by enlarging $E_0$.  This leads to the amusing question of whether Planck accelerators would
be feasible in a hypothetical case where the proton/electron masses are increased say by a 
common factor $\lambda$.\footnote{This can be achieved if all lepton, quark, and $\Lambda_{QCD}$
scales are scaled up by a common $\lambda$.  This can be done leaving gauge couplings almost
the same up to mild ``running'' (i.e. renormalization group) logarithmic changes.}

        Indeed as noted above (Sections (\ref{seca}) and (\ref{ca}) this would reduce synchrotron radiation and enhance
the rigidity of materials and of the vacuum against $e^+e^-$ pair production breakdown in
strong electric fields.  Hence taking $\lambda \gg 1$ appears to help facilitate super-Planckian
accelerators.  Clearly the cosmology of such a Gedanken universe is likely to drastically
change - the enhanced gravitational interactions may lead to quick recollapse drastically 
decreasing  by as much as a factor $\lambda^{-1}$ or even $\lambda^{-2}$ the maximal
$t_H$ (or $R_H$).  

It turns out however that physics on much shorter scales is also 
drastically modified, preventing dramatic increase in the maximal energy achieved by 
accelerators.  The simple scaling $m_e \rightarrow m_e' = \lambda m_e$ but 
$\alpha \rightarrow \alpha' \approx \alpha$ will scale down by $\lambda$ atomic and lattice
unit sizes.  From Eq. (\ref{eqaa}) above we find that the new materials will be able to withstand
much larger maximal pressure

\begin{equation}
\label{eqab}
p_{max} \propto \alpha^5 m_e^4 \rightarrow p_{max}' \approx \alpha^5 m_e'^4 \approx 
\lambda^4 p_{max}
\end{equation}

and for this reason so will be the gain in say the linear accelerator

\begin{equation}
G' = \frac{dW'}{dx'} \approx \alpha^3 {m'}^2_e = \alpha^3 m_e^2 \lambda^2 = \lambda^2 G
\end{equation}

Thus an original LINAC capable of achieving maximum energy $W_{max}= GL$, can now achieve
$W'_{max} = G'L'=\lambda W_{max}$ even if $L'$ is scaled down (with
all other dimension) by $\lambda^{-1}$.

        In general the pressures in materials due to self gravity are negligible unless
the dimensions are large.  If we have a uniform body of size $R$ (in all directions)
then the gravitational pressure is

\begin{equation}
p_G = \frac {F_{Grav.}}{Area} \approx \frac {G_N M^2 /R^2}{R^2} = \frac{G_N M^2}{R^4}
\end{equation}

If we use $\rho = GeV/a_0^3 \approx m_N/a_0^3 = m_N m_e^3 \alpha^3$ for ordinary materials
and $M=\rho R^3$ then:  $p_G = G_N \rho^2 R^2 = \frac {m_N^2 m_e^6 \alpha^6 R^2}{m_{Pl}^2}$
It exceeds $p_{max} = \alpha^5 m_e^4$ when $R \geq 10^{10} cm$ and hence hypothetical
large cold stars of such dimensions would liquify at the center.  For a given $R$ $p_G$
scales with $\lambda^8$ whereas the maximal pressure Eq. (\ref{eqab}) only with $\lambda^4$.
Hence the maximal size of new material is $R \leq (\frac {10^5}{\lambda})^2 cm$.
Thus taking $\lambda = 10^5$ (so as to allow bridging the gap between our conjectured
maximal energy $W_{max}$ (in EM accelerators) $\approx 10^{14} GeV$ and $m_{Pl}$) will make 
even $(cm)^3$ devices crush - leaving little room for Planckian accelerators.\footnote{Note
 that we do not claim that the present parameters of elementary particles are optimally
chosen so as to avoid or to facilitate Planck acceleration - though the earlier version
of this paper contained some speculation that rare Planckian collisions led to 
universes with different, lighter, Fermionic generations.  We found that Lee Smolin\cite{Smolin}
has indeed speculated that black holes do give rise after Hawking evaporation to baby
universes, and that the fundamental physical parameters are such that the rate of black
hole formation and breeding of new universes is maximized.}

\section{Summary, Comments and Speculations}

The above discussion strongly suggests that elementary particles with super-Planckian
energies may be unachievable.  Is this indicative of new physics or just a curiosity?
There is the well known example of our inability to build a Heisenberg microscope so as to
beat the uncertainty principle.  However, unlike in the celebrated case, we do not (yet!)
see a single common principle causing all our Gedanken accelerators to fail.

        It has been conjectured that Planck scale physics can manifest in the low energy regime 
by inducing effective interactions which violate all global symmetries.  An 
example is a $\frac{\lambda}{m_{Pl}} \Phi^+ \Phi \Phi^+ \Phi \Phi$ term where the
$\Phi$ bosons carry two units of lepton number.  Such a term violates $U(1) of
(B-L)$.  It endows the putative massless Goldstone
boson (Majoron), associated with a spontaneous breakdown of this Global
$U(1)$, with a finite mass.  A concrete mechanism for B-L violation involves the
formation of a black hole in a collision of, say, $\Phi^+\Phi$, followed by
the decay of the B.H. into $\Phi \Phi \Phi^+$, a final state with two units of
lepton number. In this way the violation of the global quantum numbers traces
back to the fundamental ``No Hair Theorem'' for black holes.  Exactly as in
the case of SU(5), where a virtual $X,Y$ GUTS meson can mediate nucleon decay
by
generating effective four Fermi terms, the virtual ``mini black hole''
system was conjectured to induce the $\frac{\lambda}{m_{Pl}} \Phi^+ \Phi \Phi^+
\Phi \Phi $ term.  The estimated resulting Majoron mass $M_x \approx KeV$ is
rather high.\cite{Moha}
Also Planckian black holes would constitute some irreducible environment
and may require modification of quantum mechanics

However our inability to achieve super-Planckian energies could be due to some
profound principle.  In the ``ultimate'' theory the whole super-Planckian
regime may then be altogether excluded - much in the same way that in quantum
mechanics the simultaneous definition of $x$ and $p$ to better than
$\Delta x \Delta p \leq \hbar /2$ is impossible.  All these global quantum number or
quantum mechanics violating effects will then not be there - and super Planck physics
even in terms of its indirect low energy manifestations could be completely absent.  Building
a theory of this kind is an outstanding challenge that clearly will not be attempted here.

{\bf Acknowledgments}

 We are indebted to P. Mazur for discussions which motivated the beginning of this
project. The encouragement and useful comments of Y. Aharonov, A.
Falk, O. W. Greenberg,Ted Jacobson,
J. Knight, S. Migdal, H. Neuberger, P.
Steinhardt, and E. Witten, are gratefully acknowledged.  We are
particularly indebted to W. Unruh for the
 ``Four Black Hole'' Planckian Accelerator  constituting
the most sophisticated suggestion to date, to B.
Reznik who found the   instability   which foiled the Unruh accelerator, to S. Massar
and N. Itzhaki for the ``black hole factory'' concept, to N. Itzhaki for suggesting
the ``hanging laboratory'' concept. Finally, S. Nussinov would like to acknowledge
Lev Vaidman for convincing us that the purely mechanical version of the inverted
cascade accelerator cannot work,
many discussions with Lenny Susskind, and the inspiring suggestion of S. Polyakov that
the non-existence of Planck accelerators may be a ``fourth law of thermodynamics'' and S.
Popescu for footnote(\ref{foa})).
We acknowledge the support of the Israel Science Foundation.



\newpage

\newpage
\begin{figure}
\begin{picture}(200,250)(40,0)
\vspace{5mm}
\hspace{18mm}
\mbox{\epsfxsize=120mm \epsfbox{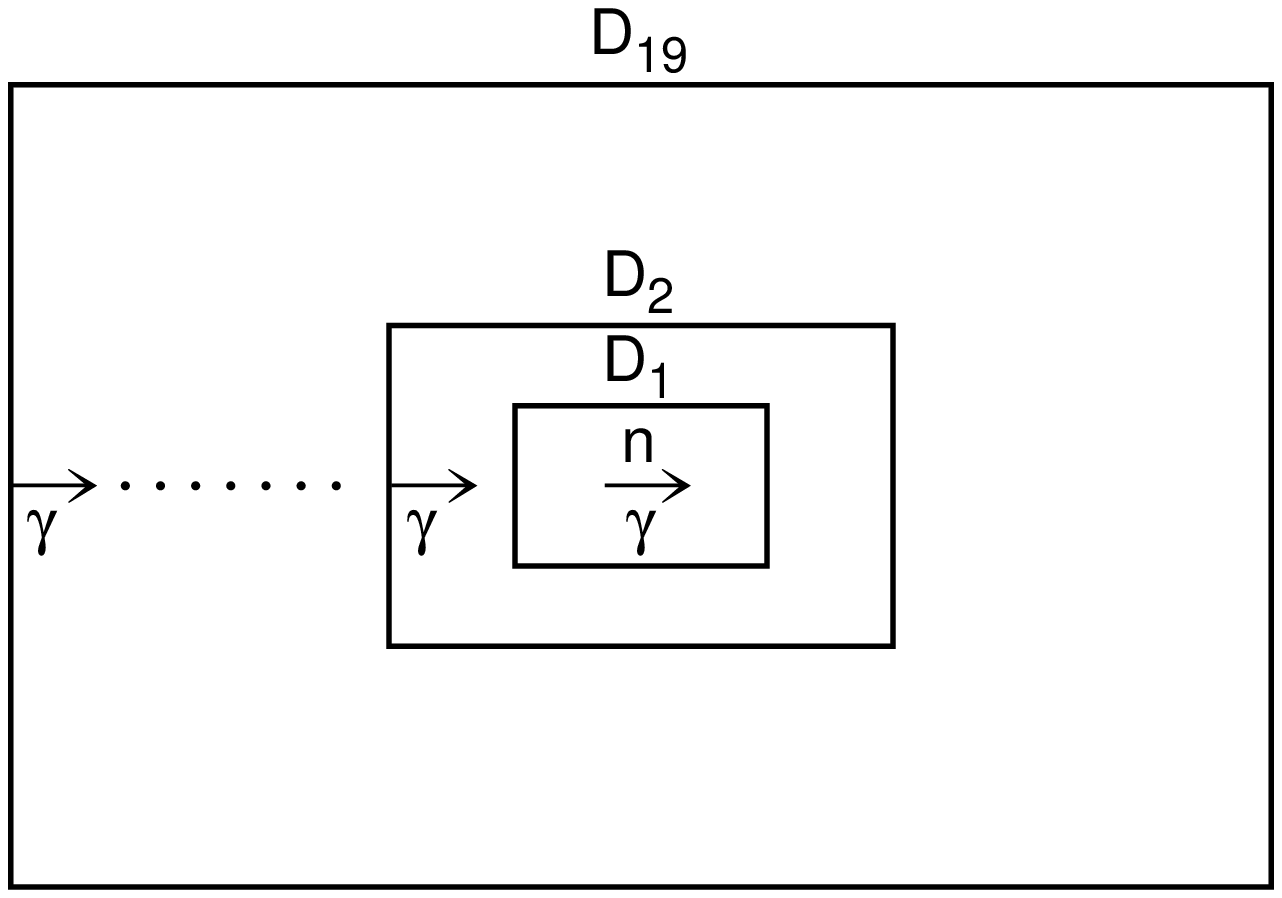}}
\end{picture}
\caption{Accelerator within accelerator within...  system designed to achieve
super-Planck energies. }
\end{figure}

\newpage
\begin{figure}
\begin{picture}(200,250)(40,0)
\vspace{5mm}
\hspace{18mm}
\mbox{\epsfxsize=120mm \epsfbox{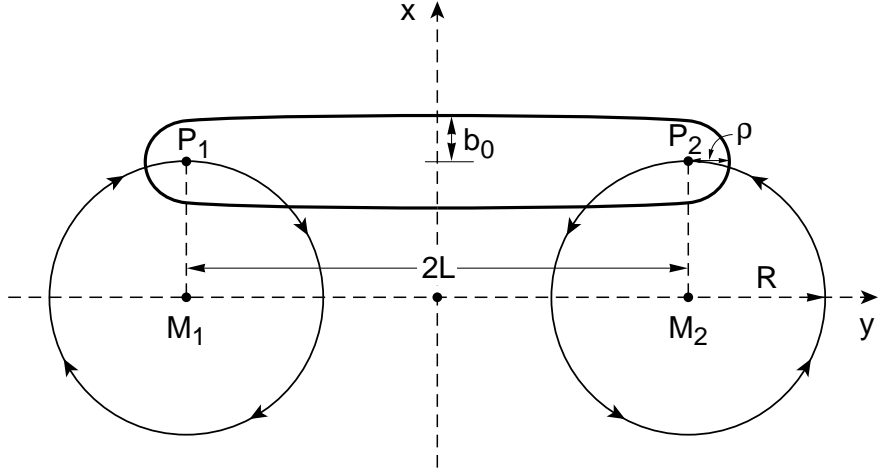}}
\end{picture}
\caption{The Unruh accelerator.  The two black holes $m_1 = m_2 = m$ go around 
the stationary more massive $M_1 = M_2 = M$ black holes in circular orbits of
radius {\it R} and in opposite directions.  The accelerating particle goes
around in the oblong ``stadium-like'' trajectory of thickness $2b_o$ with $b_o$
the impact parameter.  It gets the ``sling-shot kicks'' boosting its energy as
it goes around $P_1,\, P_2$ at times $t, t + T$ with $m_1,\, m_2$ at $P_1,\,
 P_2$ respectively.}
\end{figure}


\end{document}